\documentclass[aps,prb,groupedaddress,twocolumn,amsmath,amssymb]{revtex4}
\usepackage{amsmath}
\usepackage{graphicx}

    \usepackage{bm}

    \usepackage{dcolumn}
\newcommand\beq{\begin{equation}}
\newcommand\eeq{\end{equation}}
\newcommand\beqa{\begin{eqnarray}}
\newcommand\eeqa{\end{eqnarray}}

\begin{document}
\title{On the radial distribution function of a hard-sphere fluid}
\author{M. L\'opez de Haro}
\email{malopez@servidor.unam.mx}
\homepage{http://cie.unam.mx/xml/tc/ft/mlh/} \affiliation{Centro de
Investigaci\'{o}n en Energ\'{\i}a, U.N.A.M., Temixco, Morelos 62580,
M{e}xico}
\author{A. Santos}
\email{andres@unex.es}
\homepage{http://www.unex.es/eweb/fisteor/andres/}

\author{S. B. Yuste}
\email{santos@unex.es}
\homepage{http://www.unex.es/eweb/fisteor/santos/}
\affiliation{Departamento de F\'{\i}sica, Universidad de
Extremadura, E-06071 Badajoz, Spain}

\begin{abstract}
{Two related approaches, one fairly recent {[A. Trokhymchuk
\textit{et al.} J. Chem. Phys. {\bf 123}, 024501 (2005)]} and the
other one introduced {fifteen} years ago {[S. B. Yuste and A.
Santos, {Phys. Rev. A} {\bf 43}, 5418 (1991)]}, for the derivation
of analytical forms of the radial distribution function of a fluid
of hard spheres are compared. While they share similar starting
philosophy, the first one involves the determination of eleven
parameters while the second is a simple extension of the solution of
the Percus--Yevick equation. It is found that the {second} approach
has  a  better global accuracy and the further asset of counting
already with a successful generalization to mixtures of hard spheres
and other related systems.}
\end{abstract}

\maketitle

The fluid of hard spheres (HS) is  a key model in the use and
application of statistical mechanics to the study of the
thermodynamic and structural properties of real fluids. This is due
not only to the relative simplicity of the intermolecular potential
but also to the availability of rather accurate (albeit
semi-empirical) expressions for the corresponding equation of state
{(EOS)}\cite{EOS} as well as the exact solution of {the
Ornstein--Zernike  equation with the Percus--Yevick (PY)
closure,\cite{wertheim}} that provides an approximate analytical
expression for the radial distribution function (rdf). Yet, interest
in this system is very much alive since exact analytical expressions
both for its thermodynamic and structural properties have defied
derivation. A recent example of this interest of particular
relevance to our purpose here is the analytical expression for the
rdf {proposed} by Trokhymchuk {\em et al.}
{(TNJH)}.\cite{Trokhymchuk}

{Fifteen} years ago, two of us\cite{Bravo1} developed an analytical
method to determine  reliable expressions for the rdf $g\left(
r\right)$ of an HS fluid. Such method provides a reasonable
extension of the solution to the PY equation by imposing weak
physical requirements. Apart from being relatively simple, requiring
as the only input the {EOS} of the HS fluid, it has the further
asset of including by construction the full thermodynamic
consistence between the virial and compressibility routes to the
{EOS}. The method uses a rational function {ansatz for a function
related to} the Laplace transform of $rg\left( r\right)$, and so it
is referred to as the Rational Function Approximation (RFA) method.
It has subsequently been used to study the metastable region past
the liquid-solid transition,\cite{Bravo2} successfully adapted and
generalized to deal with other systems {(such as HS
mixtures,\cite{Bravo3} sticky hard spheres,\cite{SHS} and
square-well fluids\cite{SW})}, used to derive other structural
properties {(such as the contact values of the derivatives of the
rdf, the direct correlation function, and the bridge
function\cite{Bravo4})} and also in connection with the perturbation
theory of liquids.\cite{Robles2}

Interestingly enough, there are {several} common features in the RFA
method and in the {TNJH approach}\cite{Trokhymchuk} to derive an
approximate analytical form for $g\left( r \right)$. For instance,
the underlying philosophy is the same as evidenced through the
following text that we quote from Ref.\ \onlinecite{Trokhymchuk}:
``A third possibility towards obtaining a nonempirical analytic
representation of $g(r)$ is to assume a certain theoretically
justified functional form for $g(r)$ (e.g., a result of the solution
of the OZ equation) with adjustable parameters and impose then
certain constraints (e.g., thermodynamic consistency) to determine
these parameters.'' Such a statement is totally applicable to the
RFA method. Furthermore, both use the asymptotic behavior of the rdf
for $r \to\infty$, they {enforce} thermodynamic consistency, and
both also require as input only {a prescribed compressibility factor
$Z\equiv p/\rho k_BT$ and its associated isothermal susceptibility
$\chi=[{\partial (\rho Z)}/{\partial \rho}]^{-1}$.}
 Therefore it seems natural
to compare the outcomes of both approaches in order to assess their
merits and limitations. The purpose of this Note is to carry out
such a comparison.

In  the {TNJH approach},\cite{Trokhymchuk} the rdf is written as
\beq
g(r)=\begin{cases}
g_{\text{d}}(r),& 1<r<r_{\text{m}}\\
g_{\text{s}}(r),& r>r_{\text{m}},
\end{cases}
\label{T1}
\eeq
where $r_{\text{m}}$ ($1<r_{\text{m}}<2$) denotes the first minimum
of $g(r)$ and the ``depletion'' (d) and ``structural'' (s) parts
have the forms
\beq
g_{\text{d}}(r)=\frac{A}{r}e^{\mu(r-1)}+\frac{B}{r}\cos\left(\beta(r-1)+\gamma)\right)e^{\alpha(r-1)},
\label{T2}
\eeq
\beq
g_{\text{s}}(r)=1+\frac{C}{r}\cos\left(\omega
r+\delta\right)e^{-\kappa r},
\label{T3}
\eeq
respectively. In Eqs.\ (\ref{T1})--(\ref{T3}) and henceforth units
are taken in which the diameter  $\sigma$ has a value of 1. The
depletion and structural parts have the same functional dependence
on $r$ as the exact solution of the PY equation in the shell $1<r<2$
and in the limit $r\to\infty$, respectively. The approximation
(\ref{T1}) contains eleven parameters  to be determined as functions
of the density $\rho$ (or, equivalently, of the packing fraction
$\eta\equiv \pi\rho\sigma^3/6$). Trokhymchuk \textit{et al.} impose
five natural conditions: consistency with prescribed $Z$ and $\chi$,
continuity of $g(r)$ and its first derivative at $r=r_{\text{m}}$,
and the minimum condition at $r=r_{\text{m}}$. More explicitly,
\beq
g(1^+)=\frac{Z-1}{4\eta},\quad \int_0^\infty dr\,
r^2\left[g(r)-1\right]=\frac{\chi-1}{24\eta},
\label{T4}
\eeq
\beq
g_{\text{d}}(r_{\text{m}})=g_{\text{s}}(r_{\text{m}}),\quad
\left.\frac{\partial g_{\text{d}}(r)}{\partial
r}\right|_{r_{\text{m}}}=0,\quad\left.\frac{\partial
g_{\text{s}}(r)}{\partial r}\right|_{r_{\text{m}}}=0.
\label{T5}
\eeq
In Ref.\ \onlinecite{Trokhymchuk} the Carnahan--Starling--Kolafa
 (CSK)\cite{CSK} {EOS} is used for $Z$ and
$\chi$. Furthermore, the parameters $\gamma$ and $\mu$ are fixed at
their PY expressions,\cite{wertheim} $\kappa$ and $\omega$ being
given by parameterizations obtained by Roth \textit{et
al.}\cite{RED00} This still leaves seven parameters to be determined
under the five constraints (\ref{T4}) and (\ref{T5}). By using the
first equation of (\ref{T4}) and the first and third equations of
(\ref{T5}),  $A$, $B$, $C$, and $\delta$ are expressed in terms of
$\alpha$, $\beta$, and $r_{\text{m}}$, as well as of
$g_{\text{m}}\equiv g(r_{\text{m}})$. These remaining four unknowns
are determined by applying two conditions, namely, the second
equation of (\ref{T4}) and the  second equation of (\ref{T5}), and
``by minimizing an appropriate functional,'' the form of which is
not given in Ref.\ \onlinecite{Trokhymchuk}. In any case, due to the
highly nonlinear character of the problem, the numerical solutions
for $\alpha$, $\beta$, $r_{\text{m}}$, and $g_{\text{m}}$ are
parameterized as functions of $\eta$ in the range
$0.2\leq\rho\sigma^3\leq 0.9$.

On a different vein, in the RFA method the Laplace transform of $r
g\left(r\right) $ is taken to be given by
\begin{equation}
G\left( t\right) ={\cal L}\left\{ r g\left( r \right) \right\}
=\frac{t}{12\eta }\frac{1}{1-e^{t}\Phi \left( t\right) },
\label{1}
\end{equation}
where $$\Phi \left( t\right)
=({1+S_{1}t+S_{2}t^{2}+S_{3}t^{3}+S_{4}t^{4}})/({1+L_{1}t+L_{2}t^{2}}),$$
the  six coefficients $S_{1}$, $S_{2}$, $S_{3}$, $S_{4}$, $L_{1}$,
and $L_{2}$  {(which depend on the packing fraction)} being
evaluated in an algebraic form by imposing the following
requirements:\cite{Bravo1,Bravo2} {(i) $\chi$ must be finite and
hence the first two integral moments of the total correlation
function $ h\left( r\right) \equiv g\left( r\right) -1$, {i.e.}, $
\int_{0}^{\infty}dr\,r^nh\left( r\right)$ with $n=1,2$, must be well
defined; (ii) the approximation must be thermodynamically consistent
with a prescribed {EOS}, i.e., Eq.\ (\ref{T4}) must be verified.}
Using the first requirement one finds that $L_1$, $S_1$, $S_2$, and
$S_3$ are linear functions of $L_2$ and $S_4$. {The choice
$L_2=S_4=0$ yields the PY solution.\cite{wertheim} On the other
hand, imposing the requirement (ii) } {leads\cite{Bravo1,Bravo2} to
 explicit expressions for $ L_{2}$ and $S_{4}$
in terms of $\chi$ and $Z$. In turn, the} rdf is given by
\begin{equation}
g\left( r\right) =\frac{1}{12\eta r}\sum_{n=1}^{\infty }\varphi
_{n}\left( r-n\right) \Theta \left( r-n\right) , \label{7}
\end{equation}
with $\Theta(r) $ the Heaviside step function and $\varphi
_{n}\left( r\right) =-{\cal L}^{-1}\{ t\left[ \Phi \left( t\right)
\right] ^{-n}\}$, which can be explicitly expressed by using the
{residue} theorem.\cite{Bravo2} {To close the problem one has to
give an {EOS} to fix $Z$ and $\chi$, so all the procedure is a
function of this choice. Here, we take the same choice as in Ref.\
\onlinecite{Trokhymchuk}, namely the CSK {EOS}.\cite{CSK}}

{The form of the TNJH approximation, Eqs.\ (\ref{T1})--(\ref{T3}),
is particularly simple, while the RFA method, as happens with the PY
solution, gives an  explicit expression (\ref{7}) for each separate
shell $n<r<n+1$. On the other hand, the formal simplicity of the
TNJH is at the expense of introducing a number of unknowns exceeding
the number of constraints, so that minimization of a certain
empirical functional is needed. As a consequence, the problem
becomes highly nonlinear and the numerical solutions are fitted to
given expressions; a change in the desired {EOS} would then require
to start all over again the fitting procedure. In contrast, the RFA
only requires six coefficients which are explicitly expressed
 in terms of an arbitrary EOS. {}From a
less practical but more fundamental point of view, it can be added
that the TNJH $g(r)$ introduces an artificial second-order
discontinuity at the minimum $r=r_{\text{m}}$, while it forces
$g(r)$ to be analytic at $r=2,3,\ldots$}

\begin{figure}[ht]
\includegraphics[width=\columnwidth]{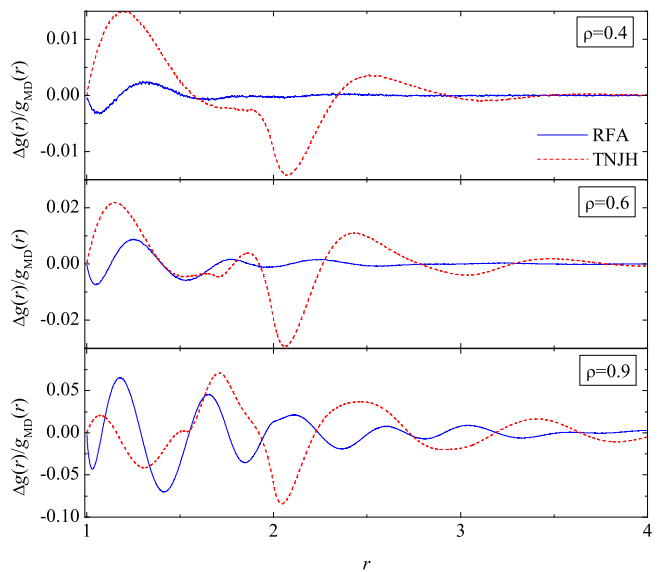}
\caption{{(Color online) {Relative deviation ${\Delta
g(r)}/{g_{\text{MD}}(r)}\equiv\left[{g(r)-g_{\text{MD}}(r)}\right]/{g_{\text{MD}}(r)}$
of the values of the rdf   of an HS fluid with respect to the
molecular dynamics  simulation data,\protect\cite{Kolafa}
$g_{\text{MD}}(r)$, as a function of $r$ for $\rho \sigma^3=0.4$
(top panel), $\rho \sigma^3=0.6$ (middle panel), and $\rho
\sigma^3=0.9$ (bottom panel). The dashed lines represent the results
of the approach of Ref.\ \onlinecite{Trokhymchuk}, while the solid
lines refer to those of the RFA method.\protect\cite{Bravo1}}
\label{fig1}}}
\end{figure}
We are now in a position to carry out the aforementioned comparison.
{Since both methods lead to reasonably accurate values for the rdf,
in Fig.\ \ref{fig1} we present the results of the relative
deviations of such values with respect to the very recent simulation
data\cite{Kolafa} for three characteristic densities.  It is quite
clear  that only in the case $\rho \sigma^3=0.9$ and for small
distances, the RFA values give a somewhat poorer agreement with the
simulation results.} In any case, and recognizing the value of the
TNJH approach,\cite{Trokhymchuk}  it is fair to say that the RFA
method {provides} a rather accurate and simple alternative to
compute the structural properties of the HS fluid. Certainly, there
are fewer parameters in the RFA method and they have a clear
physical link. Moreover,  as the authors of Ref.\
\onlinecite{Trokhymchuk} point out, the availability of analytical
results in instances where simulations may fail (such as HS mixtures
 of a large size ratio) is crucial. In this context, the fact
that we already have a simple extension of our approach to deal
successfully with these and related systems\cite{Bravo3,SHS,SW}
indicates that the RFA method is a valuable tool for the theoretical
investigation of the structural properties of {hard-core} systems.

\begin{acknowledgments}
We are grateful to J. Kolafa for kindly providing us with tables of
computer simulation results. Thanks are also due to I. Nezbeda and
A. Trokhymchuk for an exchange of electronic correspondence that
helped us to clarify some points and understand better their
formulation. The research of A.S. and S.B.Y. has been supported by
the Ministerio de Educaci\'on y Ciencia (Spain) through grant No.\
FIS2004-01399 (partially financed by FEDER funds) and by the
European Community's Human Potential Programme under contract
HPRN-CT-2002-00307, DYGLAGEMEM.
\end{acknowledgments}

\end{document}